\documentclass[twocolumn,longauthor]{aastex62}

\newcommand{\msun}{\, \mbox{M}_{\odot}}
\received{July 24, 2018}
\revised{August 11, 2018}
\accepted{August 14, 2018 by ApJ}
\shorttitle{Active Assembly of the Virgo Cluster}
\shortauthors{T. Lisker, R. Vijayaraghavan, et al.}

\begin{document}

\title{The Active Assembly of the Virgo Cluster: Indications for Recent Group Infall From Early-Type Dwarf Galaxies}

\correspondingauthor{Thorsten Lisker}
\email{thorsten@dwarfgalaxies.net}

\author{Thorsten Lisker}
\affil{Astronomisches Rechen-Institut, Zentrum f\"ur Astronomie der Universit\"at Heidelberg, M\"onchhofstra\ss e 12-14, 69120 Heidelberg, Germany}

\author{Rukmani Vijayaraghavan}
\affiliation{Department of Astronomy, University of Virginia, 530 McCormick Road, Charlottesville, VA 22904, USA}
\affiliation{NSF Astronomy \& Astrophysics Postdoctoral Fellow}

\author{Joachim Janz}
\affil{Astronomy Research Unit, University of Oulu, FI-90014 Finland}
\affil{Finnish Centre of Astronomy with ESO (FINCA), University of Turku, V\"ais\"al\"antie 20, FI-21500 Piikki\"o, Finland}

\author{John S.\ Gallagher, III}
\affil{Department of Astronomy, University of Wisconsin-Madison, 475 N.\ Charter Street, Madison, WI 53076-1582, USA}

\author{Christoph Engler}
\affil{Astronomisches Rechen-Institut, Zentrum f\"ur Astronomie der Universit\"at Heidelberg, M\"onchhofstra\ss e 12-14, 69120 Heidelberg, Germany}
\affil{Max Planck Institute for Astronomy, K\"onigstuhl 17, 69117 Heidelberg, Germany}

\author{Linda Urich}
\affil{Astronomisches Rechen-Institut, Zentrum f\"ur Astronomie der Universit\"at Heidelberg, M\"onchhofstra\ss e 12-14, 69120 Heidelberg, Germany}

\begin{abstract}
Virgo is a dynamically young galaxy cluster with substructure in its spatial and kinematic distribution. Here, we simultaneously study the phase-space distribution and the main characteristics of Virgo's galaxies, particularly its most abundant galaxy population -- the early-type dwarfs -- to understand their environmental transformation histories. Aside from known correlations with morphological types -- like the larger average clustercentric distance of late-type galaxies -- we find an intriguing behavior of early types with magnitudes $-17 \geq M_r \geq -18$. They show a large velocity spread and an asymmetric phase-space distribution, similar to the late-type galaxies and different from the early types just one magnitude brighter/fainter. Furthermore, we find a close phase-space aggregation of early-type dwarfs at large clustercentric distance and high relative velocity. Nearly all of them show signatures of disk components and their colors imply stellar ages that are  younger than the population average. They are not located closely together but spread azimuthally around the cluster center. We show that this is expected from simulations of an infalling galaxy group that slowly gets dispersed after its first pericentric passage. We thus conclude that these galaxies are recent arrivals, and that the peculiar phase-space distribution of early-type dwarfs is evidence for the ongoing growth of this galaxy population. Studying galaxies based on their phase space correlations is a unique way to compare the properties of recent and older cluster members, and to understand which environment most influenced their present-day characteristics.
\end{abstract}


%-----------------------------------------------------------------------------------
%-----------------------------------------------------------------------------------
\section{Introduction}
\label{sec:intro}

The environmental dependence of the morphological distribution of galaxies is well known. Dense environments -- clusters and massive groups of galaxies -- preferentially host early-type galaxies, while late-type galaxies are more likely to exist in low-density environments. This phenomenon has been well studied and quantified for giant ellipticals and spirals \citep{Dressler1980}, and extends to dwarf galaxies as well \citep{Binggeli1987}. Early-type dwarfs are the dominant population (by number) in the centers of massive galaxy clusters \citep{Binggeli1987}.

The presence of multiple families of early-type dwarfs has been studied by many groups, starting with \citet{Wirth84,Okamura85,Kormendy89}. While some studies interpret the early-type galaxy population from giants to dwarfs as a continuum of smoothly changing properties with decreasing stellar mass \citep[e.g.][]{GrahamGuzman2003,Chen2010,Glass2011}, early-type dwarfs are not simple scaled-down versions of bright ellipticals and lenticulars. For instance, dwarfs at stellar masses around $10^9 \, \msun$ (absolute magnitudes around $ M_r \approx -17$\,mag), have star formation histories with systematically longer timescales than those of bright early types \citep{Gavazzi2002}. The dwarfs' radial intensity profiles largely have low S\'ersic indices close to exponential, similar to late-type galaxies \citep{BinggeliCameron1991}. In the magnitude-size diagram of galaxies, \citet{Janz2008} showed that early-type dwarfs around $10^9 \, \msun$ depart from a continuous sequence of decreasing S\'ersic index from giants to dwarfs --- instead, they tend towards larger effective radii, where also the late types are found \citep{Janz2014}. Therefore, one possibility of morphologically classifying galaxies is to regard them as two parallel sequences of decreasing stellar mass: the late-type sequence, where spirals are followed by irregulars at the low-mass end, and the early-type sequence, where lenticulars of lower mass have less prominent bulges and are then followed by the bulge-less early-type dwarfs \citep{KormendyBender2012}.

Even with the similarities in  S\'ersic indices and effective radii trends, early-type galaxies are not simply descendants of late-type galaxies as we know them at present.
For instance, the internal dynamics of lenticulars show a lower stellar angular momentum and higher central concentration than that of spiral galaxies of similar masses \citep{Dressler1980,Querejeta2015}. The mere removal of gas by ram pressure and subsequent cessation of star formation from lower-mass spirals cannot explain these differences \citep{Vaghmare2015} --- one possibility to connect spirals and lenticulars in parameter space is the loss of angular momentum through mergers \citep{Querejeta2015}. Likewise, the internal dynamics of cluster early-type dwarfs are systematically different from that of present-day low-mass late types \citep{Rys2014}. Additionally, \citet{SanchezJanssen2012} pointed out that the globular cluster systems of today's low-mass late types are, on average, not as rich as those of the brighter early-type dwarfs.  Furthermore, for a sample of low-mass early-type galaxies in isolation, \citet{Janz2017} recently found that they consist of a similar mix of fast and slow rotators as those in clusters \citep{Toloba2015}. The role of these galaxies' environmental histories in affecting their internal dynamics thus remains poorly understood. 

Galaxies found in different environments today (i.e.\ cluster vs.\ group or cluster core vs.\ outskirts) have experienced a different environmental influence on their evolution for most of their lifetime \citep{Lisker2013}. Even for those galaxies that populate the red sequence of the same galaxy cluster today, the Illustris simulation \citep{Vogelsberger2014,Genel14} shows that their properties depend on their environmental history: low-mass galaxies that entered the main progenitor of the cluster at early epochs are found on the high-metallicity side of the stellar mass-metallicity relation today \citep{Engler2017,Pasquali2018}. The fact that a morphology-density relation has been found even for subtypes of early-type dwarfs \citep{Lisker2007} may be a consequence of a range of different environmental histories among this continuously growing galaxy population. 

In addition to the gradual build-up of galaxy clusters by accreting field galaxies and smaller groups, occasional accretion events of larger galaxy groups can lead to a noticeable asymmetry and disequilibrium of the spatial and dynamical distributions of cluster galaxies. Given the crossing times of more than one gigayear for clusters as massive as Virgo \citep{deVaucouleurs61,BoselliGavazzi2006}, these new members may still appear as spatially distinct subclumps (like the M49 subcluster, \citealt{Binggeli1987}) and/or phase-space aggregations after several gigayears \citep{Vijayaraghavan15,Rhee2017}, depending on their infall direction and orbit. 
{On the scale of the stellar halo of Virgo's central galaxy M87, phase-space substructures of globular clusters and planetary nebulae have already proven useful to trace the recent accretion history \citep{Romanowsky2012,Longobardi2015}.} 

Based on the distribution and motions of its galaxies, Virgo has long been known to be a dynamically young, unrelaxed cluster \citep{Huchra85,Binggeli1987}, although the X-ray intensity peak coincides with M87 \citep{Boehringer1994}. In fact, the Virgo dwarf galaxies populate the region between M87 and M84 with roughly constant number density, i.e.\ their distribution is not peaked at M87 \citep{Binggeli1987}. With M86 most likely falling in from behind \citep[e.g.][]{Zhang2015} and M87 deviating from the mean of the velocity distribution of galaxies, \citet{Binggeli1987} concluded that even the Virgo cluster core is a dynamically young structure. Moreover, using the large number of low-mass galaxies has led to confirming several previously known structures around Virgo \citep{Tully1982} as filaments that likely feed the cluster with newly infalling galaxies \citep{Kim2016}, 
{similar to the findings of \citet{Adami2009} for the Coma cluster.}

Further evidence for Virgo's active assembly comes from the velocity distribution of its early-type dwarfs.
\citet{Conselice2001a} found that the high velocity dispersion of early-type dwarfs in Virgo resembled that of late-type galaxies rather than giant ellipticals.
Therefore, \citeauthor{Conselice2001a} concluded that the Virgo cluster early-type dwarfs are a population that has been built up through infall into the cluster over time and is not a ``primordial'' cluster population.
In addition, at least a fraction of the early-type dwarfs has probably also been shaped by dwarf-dwarf mergers \citep{Paudel2017}.

Thus, we know that: the Virgo cluster's non-equilibrium dynamics indicate its active assembly, early-type dwarfs are abundant in Virgo, but we lack a complete understanding of the origin of early-type dwarfs and their environmental histories. It is therefore worthwhile to analyze the spatial and kinematic distribution of the Virgo cluster's galaxy population, especially for the abundant dwarfs, to understand their origin. Furthermore, we need to consider that high and low-mass galaxies have likely been influenced differently by various processes over their evolutionary histories, in different environments and over a range of infall histories, which in turn may have affected their morphological type, color, size, internal dynamics, and other properties. This may have led to luminosity-dependent differences in how early-type and late-type galaxies are distributed in the cluster today. 

Therefore, to understand the dynamical and environmental origin of early-type dwarfs, we investigate how galaxies in various intervals of absolute magnitude populate observer's phase space, i.e.\ how they are distributed with respect to their distance from the cluster center and their line-of-sight velocity relative to the cluster.

In Section 2, we describe the observational dataset that serves as basis for our investigation. Section 3 presents our findings of how galaxies are distributed in phase space and what their properties are. To aid the interpretation of our results, we show predictions from simulations of a group-cluster merger in Section 4. This is followed by a discussion in Section 5 and the conclusions in Section 6.

%-----------------------------------------------------------------------------------
%-----------------------------------------------------------------------------------
\section{Observational data}
\label{sec:obsdata}

Our parent sample of Virgo galaxies is based on the Virgo Cluster Catalogue \citep[VCC,][]{Binggeli1985} and was taken from \citet{Lisker2013}, comprising 713 certain and possible member galaxies down to an apparent $r$-band magnitude of $m_r \le 16$\,mag, the approximate completeness limit estimated by \citet{Weinmann2011}. Photometry was carried out on the optical imaging data release 5 of the Sloan Digital Sky Survey \citep{sdssdr5} by \citet{Lisker2007}, \citet{Janz2009}, and \citet{Meyer2014}. Details are provided in \citet{Lisker2013}.  For 59 galaxies, $r$-band magnitudes were obtained by transforming the VCC $B$-band magnitudes, as described in the appendix of \citet{Weinmann2011}. 

While the sample of \citet{Lisker2013} relied on a compilation of heliocentric velocities obtained through the NASA/IPAC Extragalactic Database (NED), a more recent compilation of velocities has been made available by the Extended Virgo Cluster Catalog \citep[EVCC,][]{Kim2014},\footnote{The EVCC also provides a catalogue of galaxies that are possibly associated to Virgo, but lie mostly at larger distance from the cluster center, beyond the VCC area. However, we intend to investigate structures that potentially indicate recent infall but have already become part of the Virgo cluster. In order to avoid any unnecessary bias, we therefore decided to restrict our sample to the VCC, since most of the additional EVCC galaxies lie beyond a projected virial radius from the Virgo center and beyond typical infall caustics in the phase-space diagram \citep{Kim2014}.} mainly from Sloan Digital Sky Survey (SDSS) spectroscopy \citep{Strauss2002}. 
The EVCC provides velocities for 668 of our 713 galaxies; the compilation from \citet{Lisker2013} adds velocities for four further objects. The galaxies with no available velocities are not grouped around any specific clustercentric distance or position in the cluster; also \citet{Kim2014} found no dependence of spectroscopic completeness on clustercentric distance.

Galaxies that are likely members of the so-called M and W clouds, located at a distance of 32\,Mpc \citep{Gavazzi1999} in the western part of Virgo, are excluded from our working sample. This is also justified because the W-M-Sheet that encompasses those clouds is unlikely to be dynamically connected to Virgo \citep{Kim2016}. Following \citet{Gavazzi1999}, our exclusion criterion is that a galaxy lies in the projected region of one of the clouds and has a velocity $v_{\rm LG}$ relative to the Local Group larger than 1900\,km\,s$^{-1}$ (with $v_{\rm LG} = v_{\rm heliocen.} + 220$\,km\,s$^{-1}$). This applies to 47 galaxies, leaving us with a working sample of 625 galaxies. Their heliocentric velocities range from $-745$ to $2845$\,km\,s$^{-1}$, with an average of $1149$\,km\,s$^{-1}$, which we adopt as the cluster systemic velocity.

We use a Virgo cluster distance of $d = 16.5$\,Mpc, corresponding to a distance modulus of $m-M = 31.09$\,mag \citep{Mei2007,Blakeslee2009}.
Wherever available, we work with individual galaxy distances from surface brightness fluctuation analyses (mostly from \citealt{Blakeslee2009}, some from \citealt{Jerjen2004} and \citealt{Tonry2001}) to convert apparent magnitudes 
to absolute magnitudes.
This is the case for 99 of our 625 galaxies; for the others we adopt $m-M = 31.09$\,mag as canonical value.

Our final working sample of 625 Virgo cluster galaxies comprises 352 objects classified as early-type galaxies (253 of these as dwarfs), 186 spiral galaxies, 54 irregulars, 19 blue compact dwarfs, and 14 galaxies intermediate between early and late type (classes ``Im/dE'' and ``Amorph''), based on the classification by \citet{SandageBinggeli1984}.
We omit those 14 intermediate objects whenever we draw comparisons between early-type and late-type galaxies. 
We consider as \emph{early-type dwarfs with disk components} those galaxies that were classified as dS0 in the VCC (see \citealt{BinggeliCameron1991} for a detailed outline of their criteria) or for which \citet{Lisker2006a} reported disk substructure from an SDSS image analysis.

%-----------------------------------------------------------------------------------
%-----------------------------------------------------------------------------------
\section{Observational findings}
\label{sec:results}

\begin{figure}
\includegraphics[width=85mm]{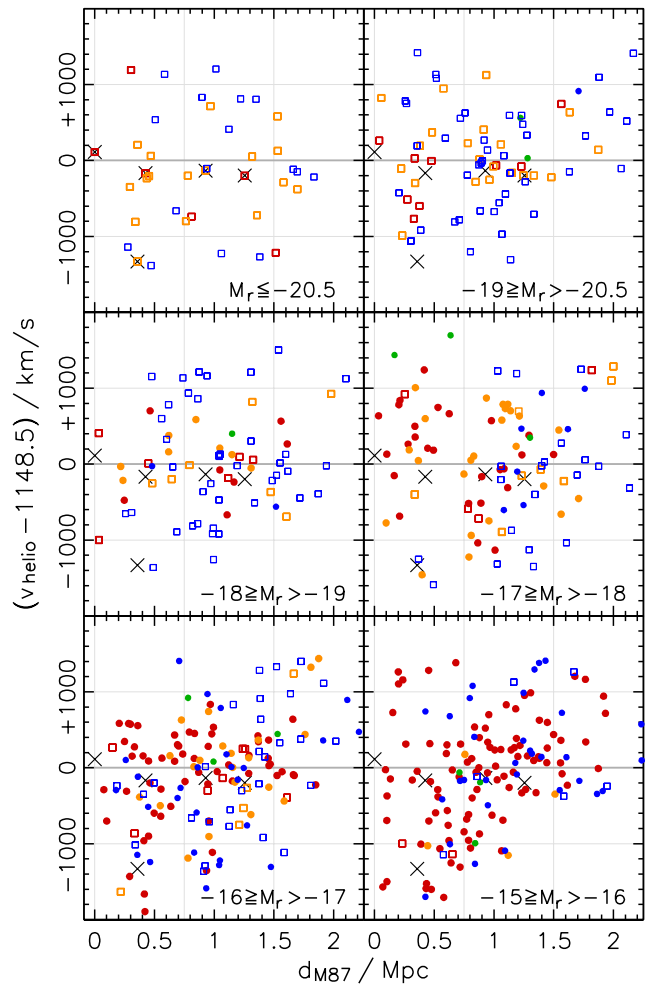}
 \caption{
Observer's phase space for our sample of 625 Virgo cluster galaxies, shown separately for different intervals of absolute $r$-band magnitude. Heliocentric velocities are relative to the sample average of $1149$\,km\,s$^{-1}$. Colors indicate the morphological class, based on the Virgo Cluster Catalogue. Red=elliptical (E/dE), orange=early-type with disk components (S0/dS0/dE(di)), green=types ``Im/dE'' and ``Amorph'', blue=spiral or irregular. Squares denote objects classified as non-dwarfs (E/S0/S), filled circles are dwarfs or irregulars (dE/dS0/Im). Black crosses denote the position of the giant early-type galaxies M87, M86, M84, M60 and M49 (left to right). For this figure, we added 40\,km\,s$^{-1}$ to the velocities of three early-type galaxies of the association at \mbox{$d_{\rm M87}$$\,\approx\,$$1.15$\,Mpc} and \mbox{$v_{\rm rel}$$\,\approx\,$$700$\,km\,s$^{-1}$} in the middle right panel, to avoid overlap with other symbols: VCC 1833 (at coordinates 1.21, 658), VCC 1896 (at 1.12, 749), and VCC 2019 (at 1.07, 748).
}
\label{fig:phasespace}
\end{figure}

The phase-space distribution of Virgo galaxies in six intervals of absolute $r$-band magnitude is displayed in Fig.~\ref{fig:phasespace}, showing heliocentric velocity (relative to the sample average of $1149$\,km\,s$^{-1}$) versus the projected distance from the central elliptical galaxy M87.
 The different morphological types of galaxies are marked by different symbols and colors: ellipticals in red, early types with disk components in orange, late-type galaxies (spirals, irregulars, blue compact dwarfs) in blue, and late/early transition types in green.
Galaxies that were classified by \citet{SandageBinggeli1984} as early-type dwarfs (their classes dE and dS0) or irregulars (Im) are denoted by filled circles, all others by open squares. The classification of the VCC was based on visual inspection of optical images; the term ``dwarf'' applied to early types that are comparatively faint and diffuse \citep{SandageBinggeli1984}. Therefore, no early-type dwarfs are found at $r$-band magnitudes brighter than $-19$ (top panels): the bright galaxy population consists of ellipticals, lenticulars, and spirals, with nearly no irregulars. At fainter magnitudes, dwarf and non-dwarf early types coexist -- classified visually by how diffuse they appear -- and likewise, both spirals and irregulars are found among the late types. In the least bright magnitude bin (bottom right panel), almost all the galaxies identified are early-type dwarfs and irregulars.

\begin{figure}
\includegraphics[width=85mm]{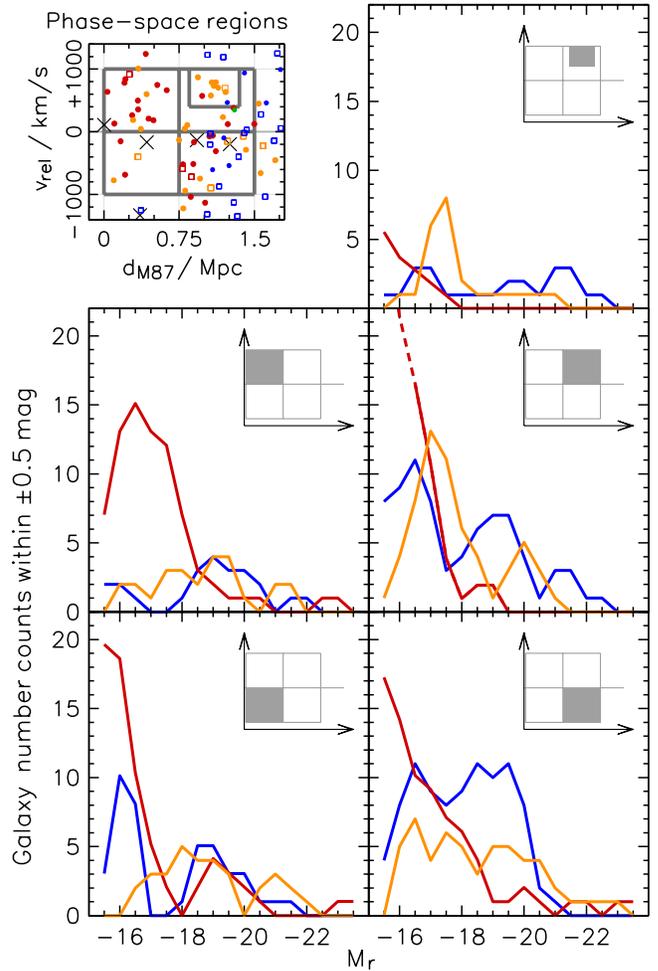}
 \caption{
   Luminosity distribution of different morphological types in various phase-space regions. Elliptical galaxies (E/dE) are displayed in red, early-type galaxies with disks in orange (S0/dS0/dE(di)), and late-type galaxies in blue (spiral/irregular classes). Each curve consists of straight lines that connect points spaced $0.5$\,mag apart; for each point we computed the galaxy number counts within $\pm 0.5$\,mag. The inset in each panel represents the phase space grid shown in the top left panel, taken from Fig.~\ref{fig:phasespace} for the magnitude interval $-17\geq M_r > -18$\,mag. For each panel, the shaded area defines the subsample of galaxies used for this respective panel. On the faint end, number counts are shown up to the completeness limit of our sample. The red curve in the middle right panel reaches a value of 29.2 on the ordinate at $-15.5$\,mag on the abscissa.
 }
\label{fig:magdistribution}
\end{figure}

The morphological and luminosity distribution of galaxies varies based on their phase-space properties. This can be seen in Fig.~\ref{fig:magdistribution}, where we display the luminosity and type distribution separately for galaxies in four distinct phase-space regions: at lower and higher clustercentric distance (i.e., $0-0.75$\,Mpc and $0.75-1.5$\,Mpc), as well as lower and higher relative velocity (0 to $\pm 1000$\,km\,s$^{-1}$ difference from the cluster average). The small insets in each panel of the figure indicate its respective phase-space region. A region encompassing an association of early-type dwarfs with disks is especially highlighted in a separate panel (top right), which is further discussed below.

We see in Fig.~\ref{fig:phasespace} that at magnitudes brighter than $-18$\,mag, galaxies with high relative velocities are mostly of late morphological type  as expected for recent infallers \citep{TullyShaya1984,Rhee2017}. The larger velocity spread of late types with respect to early types is particularly pronounced in the interval $-18\geq M_r > -19$\,mag, where the 68\%-range (16th to 84th percentile) of late type velocities is 
{1925$^{+180}_{-198}$\,km\,s$^{-1}$}
as compared to only
{912$^{+201}_{-183}$\,km\,s$^{-1}$}
for early types.\footnote{The quoted errors of these velocity ranges are the 16th and 84th percentiles from 10,000 bootstrap realizations \citep{Beers1990}.} 
Intriguingly, though, only one magnitude fainter ($-17\geq M_r >
-18$\,mag) the distribution of early types is substantially different:
their 68\%-range is now much larger, 
{1414$^{+118}_{-151}$\,km\,s$^{-1}$,}
while late types drop slightly to 
{1704$^{+358}_{-357}$\,km\,s$^{-1}$.}
Again one magnitude fainter, however, the early types' range falls
back to a lower value, with 
{1064$^{+133}_{-123}$\,km\,s$^{-1}$}
as opposed to the 
{1910$^{+194}_{-246}$\,km\,s$^{-1}$}
of the late types.

\begin{figure}
\includegraphics[width=75mm]{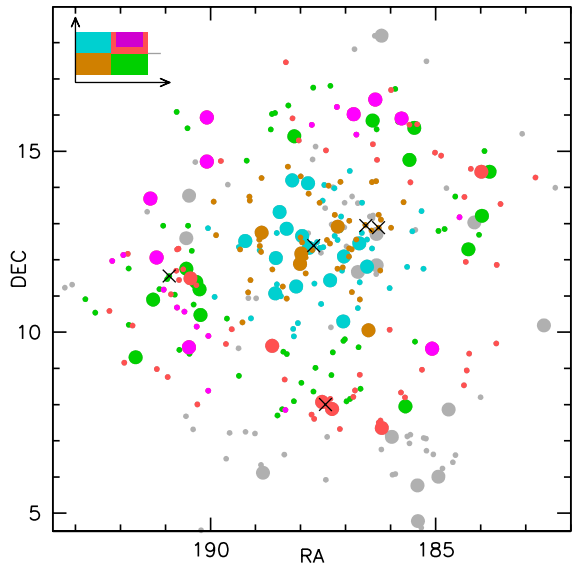}
 \caption{
Spatial distribution of Virgo cluster early-type galaxies from different phase-space regions. The inset represents phase-space, analogous to the insets in Fig.~\ref{fig:magdistribution}. Symbol colors follow the color shading of the phase-space regions; grey is used for Virgo early types outside these regions. Galaxies with $-17\geq M_r > -18$\,mag are shown with large filled circles, dots represent galaxies with fainter or brighter magnitudes. Black crosses denote the position of the giant early-type galaxies M49, M60, M87, M86, and M84 (clockwise from bottom). The galaxies shown as pink dots are a subset of the phase-space region shown by the red galaxies -- this is the group aggregated in phase-space.
}
\label{fig:radec}
\end{figure}

When going from brighter to fainter magnitudes in Figure~\ref{fig:phasespace}, the interval $-17\geq M_r > -18$\,mag is the first one in which \emph{early-type dwarfs} (filled orange and red circles) start to dominate the galaxy population in number. (Figure~\ref{fig:images} in the appendix shows SDSS images of these galaxies with $-17\geq M_r > -18$\,mag.) Even more remarkable than their large velocity distribution is the fact that they populate phase space in a rather asymmetric way: within 0.75\,Mpc, i.e.\ half the projected virial radius, the vast majority of early types has positive velocities relative to the cluster average (also see \citealt{Binggeli1987}), whereas beyond 0.75\,Mpc they are found nearly evenly on both sides.
A conspicuous association of galaxies in phase space is seen at \mbox{$d_{\rm M87}$$\,\approx\,$$1.15$\,Mpc} and \mbox{$v_{\rm rel}$$\,\approx\,$$700$\,km\,s$^{-1}$}, consisting of nine early types, eight of which show signs of a stellar disk component. \emph{As Figs.~\ref{fig:phasespace} and \ref{fig:magdistribution} illustrate, this is the only phase-space region where \emph{early-type dwarfs with disks} (filled orange circles) constitute the dominant population at these luminosities.} 

The morphological mix of this phase-space region differs significantly from that of the region with ``opposite'' velocity. Compare the middle right and bottom right panels in Fig.~\ref{fig:magdistribution}: both are regions with radius from $0.75-1.5$\,Mpc but the middle right panel has positive radial velocity and the bottom right panel has negative velocity relative to the cluster average. The cumulative binomial probability of drawing as many or more early types with disks from the latter as we find in the former is 0.4\% in the $-17\geq M_r > -18$\,mag interval.

Even when disregarding whether or not an early-type has a disk
component according to the literature, the phase-space distribution of
early types (red and orange) seen for the $-17\geq M_r > -18$\,mag
interval cannot be explained as a mere random occurrence. When
focusing again on clustercentric distances between 0.75\,Mpc and
1.50\,Mpc, 59\% of the early-type population in this magnitude range
-- 19 out of 32 galaxies -- fall \emph{outside} of a low velocity
corridor of $\pm500$\,km\,s$^{-1}$ around the cluster average. In
contrast, the fraction is only 30\% for all other 108 faint early
types (with $M_r > -19$\,mag but excluding the said magnitude range)
at those clustercentric distances. When randomly drawing 32 data
points from the 108 other faint early types 
{100,000 times (without replacement), a fraction of 59\% is reached in less than 0.01\% of cases and even 50\% is reached only in 0.3\% of cases.} 
In comparison, the fraction of faint \emph{late-type} galaxies ($M_r > -19$\,mag) outside the low velocity corridor is 58\% in the same clustercentric distance range, consistent with being an infalling population. This phase-space structure is reflective of the Virgo cluster's overall complex, non-relaxed dynamical state.

The well-known morphology-density relation of galaxies \citep[e.g.][]{Binggeli1987} can be seen clearly in Fig.~\ref{fig:magdistribution}: the overall fraction of late-type galaxies is higher at larger clustercentric distances (right panels). Additionally, it is striking how galaxy type fractions vary noticeably with the phase-space locus, including the variation in the fraction of late-type galaxies of all magnitudes (i.e.\ the varying shape and amplitude of the blue line in each phase-space region), the relative fraction of intermediate-magnitude ($-18$\,mag) early-type galaxies, as well as the relative number of early-type galaxies fainter than $-16$\,mag. The luminosity distribution of ellipticals (red curves in Fig.~\ref{fig:magdistribution}) differs significantly between the phase-space regions that are illustrated in the middle left and bottom left panel of the figure, and also between the middle left and middle right panel (K-S test probability 0.2\% and 0.0\%, respectively, for being drawn from the same parent distribution).

The spatial distribution of the early-type galaxies of the various phase-space regions is shown in Fig.~\ref{fig:radec}. It illustrates that none of the phase-space groups is clustered in a narrow spatial region, but that each of them contributes galaxies isotropically around the cluster center, M87. \emph{This is particularly relevant for the close phase-space association of early-type dwarfs with disks: these galaxies are spread out over the cluster, and are thus not a bound group of objects that just entered the cluster.} Simulations of group accretion are necessary to understand how such a peculiar phase-space structure can occur, as we see in the following section.

\begin{figure}
\includegraphics[width=85mm]{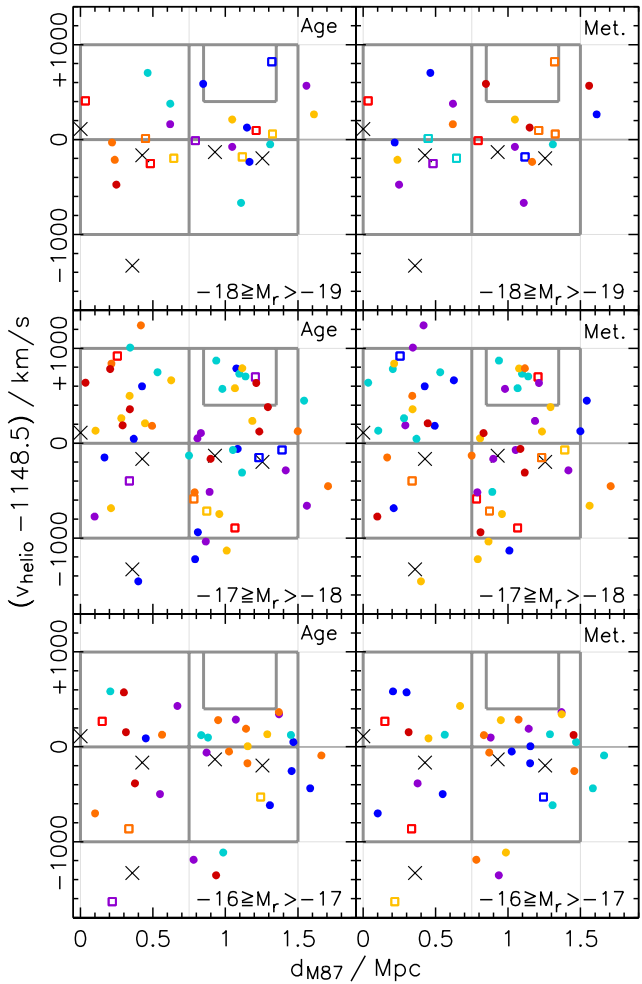}
\caption{
  Observer's phase space for 119 faint early-type galaxies of the Virgo cluster from \citet{Urich2017}, shown separately for different intervals of absolute $r$-band magnitude and for age-sensitive (left) and metallicity-sensitive color (right), based on optical\,--\,near-infrared photometry. The symbol color of a galaxy indicates whether its color at the effective radius is bluer or redder than the red sequence at its $r$-band magnitude: the bluest 20 galaxies are shown with purple symbols, the next 20 with blue symbols, then follow cyan, yellow, and orange (20 galaxies each), and the reddest 19 galaxies are shown with red symbols. Symbol shapes are as in Fig.~\ref{fig:phasespace}. Small velocity offsets were added to the same three galaxies as in Fig.~\ref{fig:phasespace} to avoid overlapping symbols. Grey boxes denote the different phase-space regions as in Fig.~\ref{fig:magdistribution}.
}
\label{fig:urich}
\end{figure}

Given the known stellar population diversity of Virgo cluster early-type dwarfs \citep{Michielsen2008,Kim2010}, we examine whether there are systematic differences between the phase-space regions. The recent optical\,--\,near-infrared photometric study of \citet{Urich2017} provides us with colors and color gradients for 119 Virgo early-type dwarfs. \citeauthor{Urich2017} defined an age-sensitive and a metallicity-sensitive color by combining $g$, $r$, $i$, and $H$ bands into linear combinations that lead to a nearly orthogonal age-metallicity grid for simple stellar populations. In the left-hand panels of Figure~\ref{fig:urich} we color-code the early-type dwarfs in six steps of age-sensitive color, from purple (youngest 20 galaxies) over blue, cyan, yellow, and orange (20 galaxies each) to red (oldest 19 galaxies). The right-hand panels analogously present metallicity-sensitive color from purple (lowest) to red (highest).

We can see that the peculiar phase-space aggregation in the interval $-17\geq M_r > -18$\,mag hosts only a single galaxy with very old stars; several galaxies have comparatively young stars. While this would seem to go hand in hand with the high fraction of disk features in this set of early-type dwarfs \citep[cf.][]{Paudel2010a}, it is not unique when compared with other phase-space regions: depending on where one places their boundaries, a variety of different stellar population mixes can be found. 
{For instance, dwarfs in the $-17\geq M_r > -18$\,mag interval that are close to the giant early-type galaxies in phase space have mostly blue age-sensitive colors, whereas dwarfs at low clustercentric distances that have positive radial velocities have mostly red colors. Relating the galaxy colors to clustercentric distance alone would thus }
not show the entire picture, as the colors also vary with velocity at a given distance.

%-----------------------------------------------------------------------------------
%-----------------------------------------------------------------------------------
\section{Predicted phase-space signatures of recent group accretion}
\label{sec:sim}

\begin{figure*}
\includegraphics[width=\textwidth]{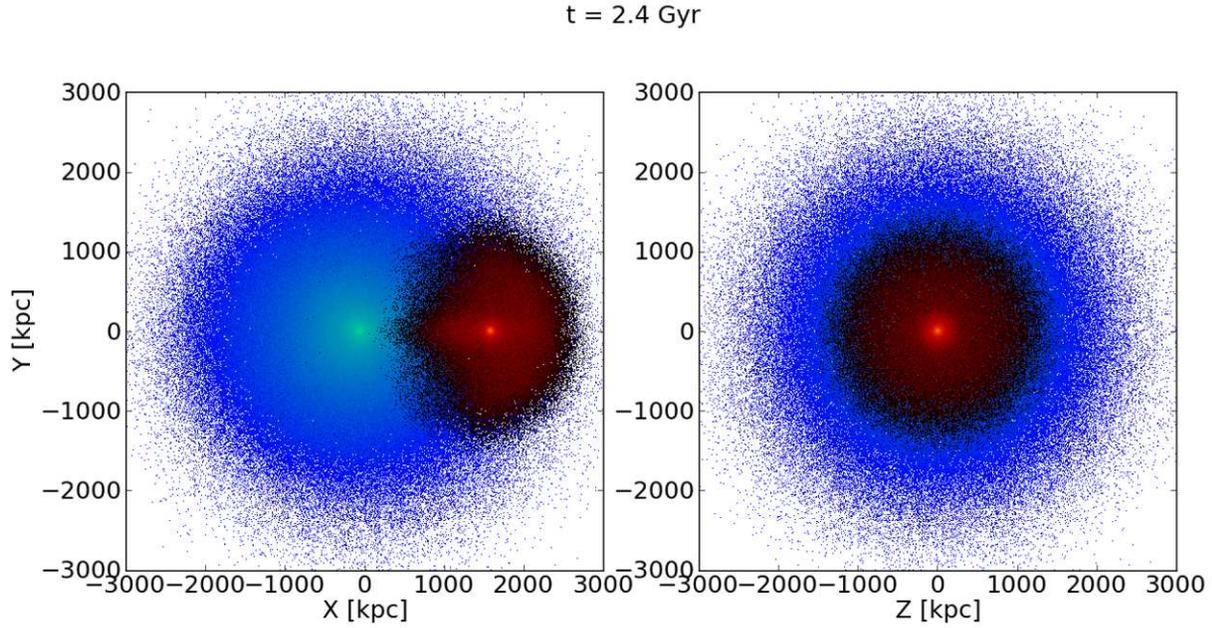}
 \caption{View of the merger perpendicular to the direction of the merger (left) and parallel to the merger line of sight (right). The cluster's particles are colored in blue-green and the group's particles are in red.}
\label{fig:gcmerger}
\end{figure*}

\begin{figure}
\includegraphics[width=85mm]{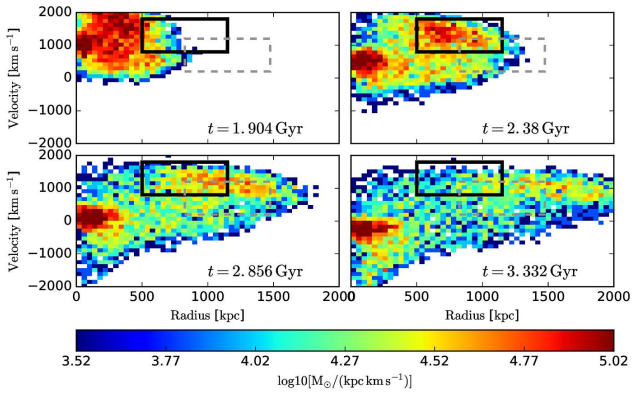}
 \caption{Phase space distribution of a group that has fallen in parallel to the observer's line of sight. The grey dashed boxes in each panel correspond to Box 1 in the later Virgo figures. The black boxes have the same velocity and radial distances as the grey boxes, but are positioned such that they coincide with the bulk of the group halo at apocentric passage, 2.4 Gyr. Pericentric passage is at 1.2 Gyr. Masses are calculated assuming a particle mass of $10^8 \mbox{M}_{\odot}$. This figure shows a $5 \times 10^{13} \, \mbox{M}_{\odot}$ group falling into a  $5 \times 10^{14}\, \mbox{M}_{\odot}$ cluster.}
\label{fig:group_phasespace}
\end{figure}

\begin{figure}
\includegraphics[width=85mm]{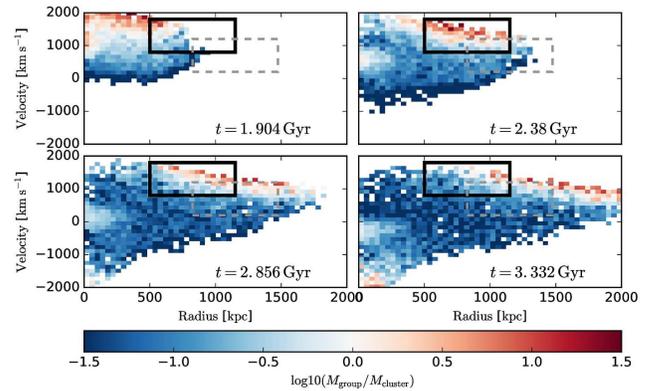}
 \caption{$\log_{\rm 10} (M_{\rm group} / M_{\rm cluster})$ in phase space, for the same group-cluster merger. The black boxes have the same velocity and radial distances as the grey boxes, but are positioned such that they coincide with the bulk of the group halo at apocentric passage, 2.4 Gyr. Pericentric passage is at 1.2 Gyr.  This plot should tell us if the galaxies in each box are primarily group galaxies or cluster galaxies.  }
\label{fig:ratio_phasespace}
\end{figure}

\begin{figure}
\includegraphics[width=85mm]{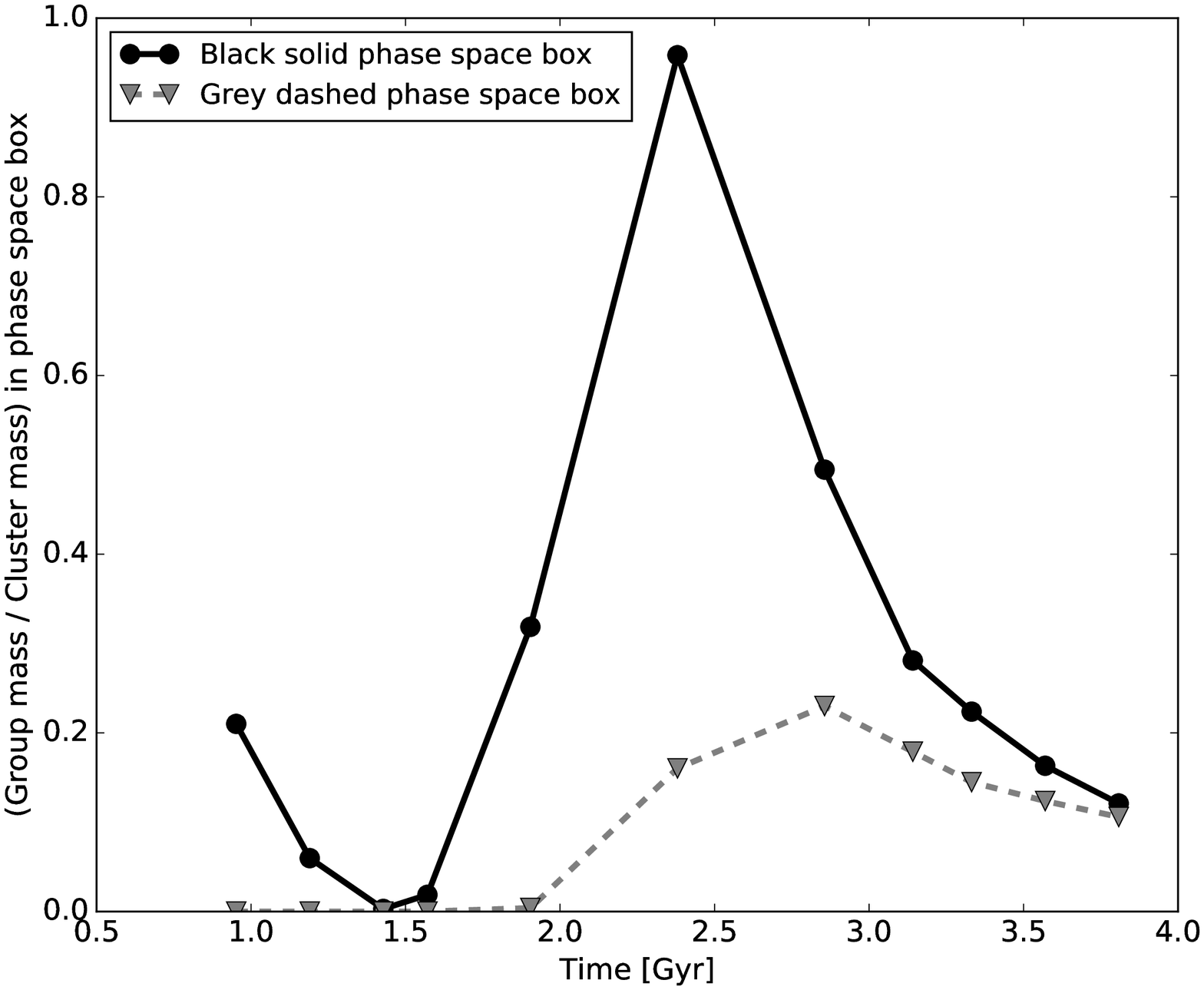}
 \caption{$(M_{\rm group} / M_{\rm cluster})$ averaged over the black box and grey dotted box in  Figures~\ref{fig:group_phasespace} and ~\ref{fig:ratio_phasespace} as a function of time.}
\label{fig:groupcluster_ratio_box}
\end{figure}

The association of galaxies in phase space clustered near \mbox{$d_{\rm M87}$$\,\approx\,$$1.15$\,Mpc} and \mbox{$v_{\rm rel}$$\,\approx\,$$700$\,km\,s$^{-1}$} likely corresponds to the remnants of the outer halo of an infalling group along the line of sight. We examine this possibility using $N$-body simulations of  idealized cluster mergers \citep[from ][hereafter V15]{Vijayaraghavan15}. This model does not have resolved galaxies, and randomly chosen particles in the $N$-body simulation are tagged with galaxy models. The results here are averaged over 100 such random realizations of cluster galaxies. Dynamical friction is also not explicitly accounted for, as further justified in V15. V15 quantified the phase-space properties of an infalling group's galaxies during its merger process, and showed that an infalling group, during a head-on collision,  is dynamically dissociated into two distinct components: the dense core, that oscillates around the cluster center while being tidally disrupted, and a diffuse outer `halo' that expands outwards after the group's core passage. 
Figure~\ref{fig:gcmerger} shows a snapshot of the merger viewed perpendicular to the merger direction -- where the group's and cluster's distinct cores are visible, and parallel to the merger -- where the cores are superimposed.  The core and halo components, when the merger is parallel to the line of sight, are visible in phase space as distinct components.

These properties are illustrated in Figure~\ref{fig:group_phasespace}, where we show the distribution of the infalling group in phase space. Here, the  one dimensional velocity along the $y$-axis is the velocity along the line of sight parallel to the infall direction with respect to the center of mass of the group-cluster system, and the radial distance along the $x$-axis is the projected radial distance from the center of mass.  The initial mass of the cluster is $5 \times 10^{14} \mbox{M}_{\odot}$ and the group is $5 \times 10^{13} \mbox{M}_{\odot}$. The panels in Figure~\ref{fig:group_phasespace} show the group's phase-space distribution from $t = 1.9$ Gyr (after the group makes its first core passage) to $t = 3.3$ Gyr. During the group's first apocentric passage, from $t \simeq 2.2$ to $3$ Gyr,  the group's core dynamically separates from the outer halo in phase space.  The core remains close to the center, with zero radial separation, but moves on the $y$-axis as seen by the variation in radial velocity at different timesteps. The outer halo remains at large radial distances, and piles up in an overdense shell/ring near the orbital apocenter. In phase space,  this feature manifests itself as an overdensity at large radii, with high radial velocities. This feature is highlighted by the black box at $t = 2.38$ Myr; the same box is drawn in the other timesteps for comparison. The grey dashed box, of the same size as the black box, is centered on the corresponding association of galaxies in Virgo near \mbox{$d_{\rm M87}$$\,\approx\,$$1.15$\,Mpc} and \mbox{$v_{\rm rel}$$\,\approx\,$$700$\,km\,s$^{-1}$} 

The phase space region highlighted by the black box is a distinct signature of recent group infall along the line of sight. V15 find similar signatures across a range of merger mass ratios and over a range of viewing angles, from along the merger to $\sim 45 \deg$ to the merger. The group core is not  prominent in phase space, especially against the existing cluster. Figure~\ref{fig:ratio_phasespace} shows the distribution of $\log_{\rm 10} (M_{\rm group}/M_{\rm cluster})$: we see a significant overdensity of group galaxies within the black-boxed halo phase-space region at $t \simeq 2.3 - 2.9$ Gyr, but there is no corresponding overdensity of the group core with respect to the cluster core. The group halo phase-space region, which forms a ring at a radius of $\sim 500 - 1200$ kpc, with peculiar velocities of $800 - 1800$\,km s$^{-1}$, is dominated by group galaxies over cluster galaxies by a mass fraction of up to $10:1$ in parts of the box. Figure~\ref{fig:groupcluster_ratio_box} shows the time evolution of the total group mass divided by the total cluster mass within these boxes. We clearly see the spike in group mass fraction at $t = 2 - 3$ Gyr.  These galaxies do not pile up as a spatially distinct compact subcluster, but are azimuthally distributed about the cluster center within the radial range encompassed by the box. \emph{Such an overdensity identified in phase space, without any obvious spatial clustering, is a likely signature of an infalling group along the line of sight. }

Of course this phase-space identification is not the only or the most precise identifier of infalling groups,
also because it cannot identify the group core, which is superimposed on the cluster core. The hierarchical clustering technique of \citet{Serna96} can alternatively be used to identify infalling groups, as by \citet{Adami2005}. This technique is particularly useful in finding gravitationally bound groups by using phase space (two spatial dimensions and one radial velocity dimension) within clusters. \citet{Adami2005} used this technique to identify 17 infalling subgroups in the Coma cluster, including two major subgroups. In contrast, our method illustrated here is a method to find the \emph{remnants} of infalling groups that \emph{do not necessarily remain gravitationally bound} once they enter the cluster
-- the quickly stripped galaxies in the infalling group's outskirts. Therefore the V15 numerical simulation here is not a precise simulation of the Virgo cluster, but illustrates the phase-space properties of the outskirts of an infalling group.

%-----------------------------------------------------------------------------------
%-----------------------------------------------------------------------------------
\section{Discussion}
\label{sec:discuss}

\subsection{Signatures of active assembly} 

Galaxy populations that get accreted by the cluster move through different phase-space regions over time, as demonstrated by \citet{Rhee2017}: most galaxies that are now experiencing their first infall are still at projected clustercentric distances beyond one virial radius, whereas galaxies around their first pericenter passage are obviously found at low distances from the center but with high clustercentric relative velocities. Galaxies that already fell into the cluster many gigayears ago populate regions of lower relative velocity, and follow a virialized, close to Gaussian velocity distribution.
Therefore, a broad relation of infall epoch with both velocity and clustercentric distance emerges (as well as with local density, see \citealt{Lisker2013}). 

This dependence of galaxies' phase-space distribution on their infall histories can shed further light on the origin of the morphology-density relation of cluster galaxies \citep{Dressler1980,Lisker2007}. \emph{If} the cluster environment continually transforms galaxies in their (optical) appearance and structure, this leads to a relation of transformational progress with clustercentric distance, velocity, and density. Alternatively, a different perspective on the morphology-density relation is that the evolution of the baryonic structure of galaxies, including the build-up of their stellar populations, has been systematically different for early infallers than for recently accreted galaxies
 {(see, e.g.\ \citealt{Pasquali2018}).} 
The latter have likely spent more time in groups, i.e.\ in parent haloes of lower mass and lower velocity dispersion \citep{Lisker2013}. Thus the processes that governed late-infall galaxies' gas accretion, star formation activity, and interaction rates were likely systematically different from those of early infallers. This ``history bias'' \citep{DeLucia2012b} unavoidably plays a role in influencing cluster galaxy populations \citep[see][]{DeCarvalho2017} --- and the galaxies' properties hold the key to distinguishing between the different infallers as well as the high-velocity tail of the older component of the cluster.

In this sample of Virgo galaxies we observe a different morphological mix for different phase-space regions (cf.\ Fig.~\ref{fig:magdistribution}). Motivated by this, we further examine the processes resulting in the transformation of infalling galaxies. First, since morphological types have traditionally been defined at optical wavelengths, the mere cessation of star formation can already be sufficient to ``transform'' an irregular galaxy into an early-type dwarf or a spiral into a lenticular.\footnote{This is \emph{not} saying that these galaxy types exhibit the same structural and dynamical properties -- there are, in fact, systematic differences \citep{Grebel2003} -- but it is merely a statement about visual galaxy classification.} The gas removal necessary to halt star formation may be caused by the ram pressure of the hot intra-cluster medium (ICM, \citealt{Schindler1999}), which can strip a galaxy fully or partially of its disk gas \citep{GunnGott1972,Chung2009}. Once this primary gas reservoir is gone, it takes about a gigayear for the blue, massive stars to die and thus for the stellar population to redden significantly \citep{Boselli2008a}. 
Additionally, the stripping of a galaxy's surrounding gaseous halo -- which is much less bound than the disk gas -- could have set in already much earlier, when the galaxy was still in a group environment. While this ``starvation'' will only slowly affect the star formation rate over the course of several gigayears \citep{Boselli2008a,Fillingham2016}, it could have caused at least some galaxies' reddening and transformation, well before they arrived to their present-day cluster environment.

Transformation of galaxies also refers to changes in a galaxy's stellar distribution and internal dynamics, induced by environmentally caused changes in the galaxy's mass distribution. While ram pressure stripping could cause such a change in specific cases when large amounts of gas are removed \citep{SmithRory2012}, more common mechanisms that cause such changes are the tidal interactions with other galaxies or with the overall cluster potential. However, as shown by the simulations of \citet{Bialas2015} and \citet{Smith2015}, such transformational processes -- like the thickening of the stellar structure of low-mass galaxies, and decrease in rotational support -- is only significant on specific orbits close to the cluster core: these processes do not induce significant structural or dynamical transformation on most typical cluster orbits. 

Since the net efficiency of external tidal forces is sensitive to how a galaxy moves in phase space and how long it has already experienced the influence of the cluster potential, tidal transformation processes also contribute to any correlation between galaxy properties with phase-space locus. At a given density, galaxy interaction rates are proportional to the velocity dispersion: galaxies in more massive clusters with higher velocity dispersions undergo more frequent galaxy-galaxy interactions than those in lower mass groups. These high speed fly-by interactions do little damage; for two interacting or colliding galaxies to merge, their relative velocities must be comparable to their internal velocity dispersions \citep{Tremaine81}. Galaxy-galaxy mergers are more common in low-mass groups \citep{Vijayaraghavan13}.  Therefore, galaxies that fall in as parts of groups can undergo pre-processing that transforms their internal dynamics, and this can be seen in the regions of phase space occupied by group remnants.

More massive, luminous galaxies have deeper gravitational potential wells, and are thus less prone to the loss of gas and/or stars due to the above mechanisms. In general, there is a mass dependence of several fundamental scaling relations that reflect a combination of underlying physical processes. From giants towards dwarfs, the Faber-Jackson relation between the luminosity and velocity dispersion of early-type galaxies changes its slope \citep{Matkovic2005}, the relation of galaxy luminosity and size -- using stellar half-light radius -- broadens significantly \citep[e.g.][]{Janz2008}, the stellar rotational support increases and then appears to decrease again (compare \citealt{Emsellem2011} with \citealt{Guerou2015}), and the efficiency of converting baryons into stars peaks and turns around \citep[e.g.][]{Behroozi2013}.

Different galaxy groups can contribute different ratios of bright, intermediate, and faint galaxies when accreted onto a cluster, since groups show a variety of faint-end slopes of the galaxy luminosity function, as well as other diverse properties \citep{TullyTrentham2008}. Also, in massive groups with non-Gaussian velocity distributions, bright and faint galaxies are not in dynamical equilibrium \citep{DeCarvalho2017}. An additional aspect that could affect the galaxy composition of the accreted group is the sweeping of field galaxies shortly before cluster infall.  \citet{Vijayaraghavan13} find that when smaller groups merge with clusters, they can `sweep-up' surrounding galaxies. These infalling galaxies would have otherwise fallen in as field galaxies, but now spend a brief period in the group, where they can be stripped or interact with other galaxies. After cluster infall, these recently accreted group galaxies, which are the most loosely bound to their group, are rapidly removed from the group's gravitational influence; they do retain their group infall velocities. 

All of the above aspects help us understand how the phase-space of a dynamically young cluster like Virgo can be populated so differently by its galaxies in different magnitude intervals (cf.~Fig.~\ref{fig:phasespace}). This phase-space population reflects the contributions that various accreted galaxy groups made to the cluster population. Low-mass galaxies, which are the most numerous, are therefore particularly good tracers of cluster infall and accretion
{ (see, e.g.\ \citealt{Adami2009}).} 
Furthermore, since their dynamical friction timescales are much higher than cluster dynamical times, their kinematic signatures are preserved over a longer time than that of the giants \citep{Conselice2001a}. Given the different pre- and post-processing scenarios  \citep{Vijayaraghavan13} that the galaxies may have experienced depending on the specific conditions in their former host group, the result is a phase-space diversity of cluster galaxy properties. We thus regard our observations as evidence for the continuous and active assembly of Virgo.

\subsection{Remnants of an infalling group in phase space}

A particularly interesting piece of evidence for recent and ongoing infall in Virgo is the conspicuous phase-space aggregation of several early-type dwarfs with disk components (see Sect.~\ref{sec:results}). This aggregation is consistent with the predictions from simulations for a group that fell into the cluster close to our line of sight (see Sect.~\ref{sec:sim}; V15): those galaxies that formerly belonged to the group's outskirts are spatially spread out over the cluster but concentrated in a narrow phase-space region. According to the simulations, the remnant of the group core should be found at low clustercentric distance, i.e.\ it is ``hidden'' among the cluster galaxies in that phase-space region. 

Even if we do not consider the former group core, the fact that the former group outskirts -- now found in the peculiar phase-space aggregation -- contribute 10 of the 105 galaxies in the interval $-17\geq M_r > -18$\,mag would imply that the accreted group must have been rather massive. This seems roughly consistent with our assumption in Sect.~\ref{sec:sim} of a group with 10\% of the total cluster mass. Any attempt to infer the former group mass more accurately from the observed galaxies would, however, be rather unreliable: firstly, it is difficult to predict exactly what fraction of the phase-space aggregation was \emph{not} brought in by the accretion event (cf.\ Fig.~\ref{fig:groupcluster_ratio_box}), and secondly, the luminosity distribution of galaxies differs significantly between the different phase-space regions (see Fig.~\ref{fig:magdistribution}), so it is not straightforward to extrapolate from this one magnitude interval to the entire population.

The question still remains as to why this phase-space aggregation is prominent in the $-17\geq M_r > -18$\,mag bin (the first to be dominated by early-type dwarfs, and the peak of the early-type dwarfs with disks population), but not in any of the other luminosity bins. Of course at the more luminous end small number statistics make it more difficult to identify any significant aggregations. The aggregation in $-17\geq M_r > -18$ stands out because of the similarity in galaxy type -- early types with disks -- and identical radial and velocity properties without any spatial correlation.  The effect is less prominent when shifting the magnitude bins by $0.2-0.5$\,mag, indicating that this $-17\geq M_r > -18$ bin may be a special transition zone from late-type to early-type for infalling galaxies. If these galaxies are members of an infalling group's outskirts,  possibly accreted just before cluster infall, they have likely lost their gas in the group or soon after cluster infall and have recently stopped star formation but still `remember' their infall dynamics. Thus we could be witnessing a special population that is undergoing morphological transformation \emph{and} retains infall dynamical histories. Further investigations of these galaxies' internal dynamics are being carried out with MUSE spectroscopic measurements to help us understand their dynamical and transformational history.

Also, the presence of the recently accreted group's core in the central Virgo region could be one aspect in understanding why the Virgo cluster core itself is not dynamically relaxed: as noted by \citet{Binggeli1987}, the Virgo cluster core shows an intriguingly complex structure and motion. Despite coinciding with the peak of the X-ray emission, the central elliptical M87 is 150\,km\,s$^{-1}$ off of the cluster's mean velocity and also does not sit at the center of the galaxy number density distribution of ``cluster A'', the main component of Virgo. Moreover, \citet{Binggeli1987} found that the early-type dwarfs in a 2-degree circle (0.6\,Mpc) around M87 have an average heliocentric velocity that is systematically higher than the cluster mean. This is clearly seen in our Fig.~\ref{fig:phasespace} for the interval $-17\geq M_r > -18$\,mag (which is the magnitude range with most early-type dwarf velocity measurements at the time of \citealt{Binggeli1987}) and could thus be an indication for the ``hidden'' accreted group core.

An ongoing merger close to the line of sight had already been suggested by \citet{Binggeli1993} to explain the asymmetric velocity distribution of early-type dwarfs in the Virgo cluster center. This interpretation was based on the above positive relative velocity of M87 and nearby dwarfs, and the strongly negative velocities of M86 and some galaxies around it \citep{Markarian1961,Binggeli1987}. In the scenario outlined by \citet{Binggeli1993}, the M86 subclump would be several times less massive than the M87 subclump, which, at first glance, seems to fit our above picture of a recent infall of a massive group. However, the velocity difference of 2000\,km\,s$^{-1}$ between M86 and the phase-space aggregation is very large and they move in opposite directions. In our simulations (Sect.~\ref{sec:sim}), the former group's core does oscillate around the cluster center, but this oscillation is damped rather quickly after the initial passage, making it unlikely that M86 was the core of the very same infalling group of which the phase-space aggregation possibly represents the former outskirts.
Nonetheless, \citet{TullyShaya1984} concluded that Virgo may currently be experiencing a merger with a ``cloud of spiral galaxies'' that began about 4 Gyr ago. Although the connection to our above case remains unclear, we note that the phase-space aggregation comprises a number of bright spiral galaxies, more than the other phase-space regions (see Fig.~\ref{fig:magdistribution}).

\begin{figure*}
\includegraphics[width=\textwidth]{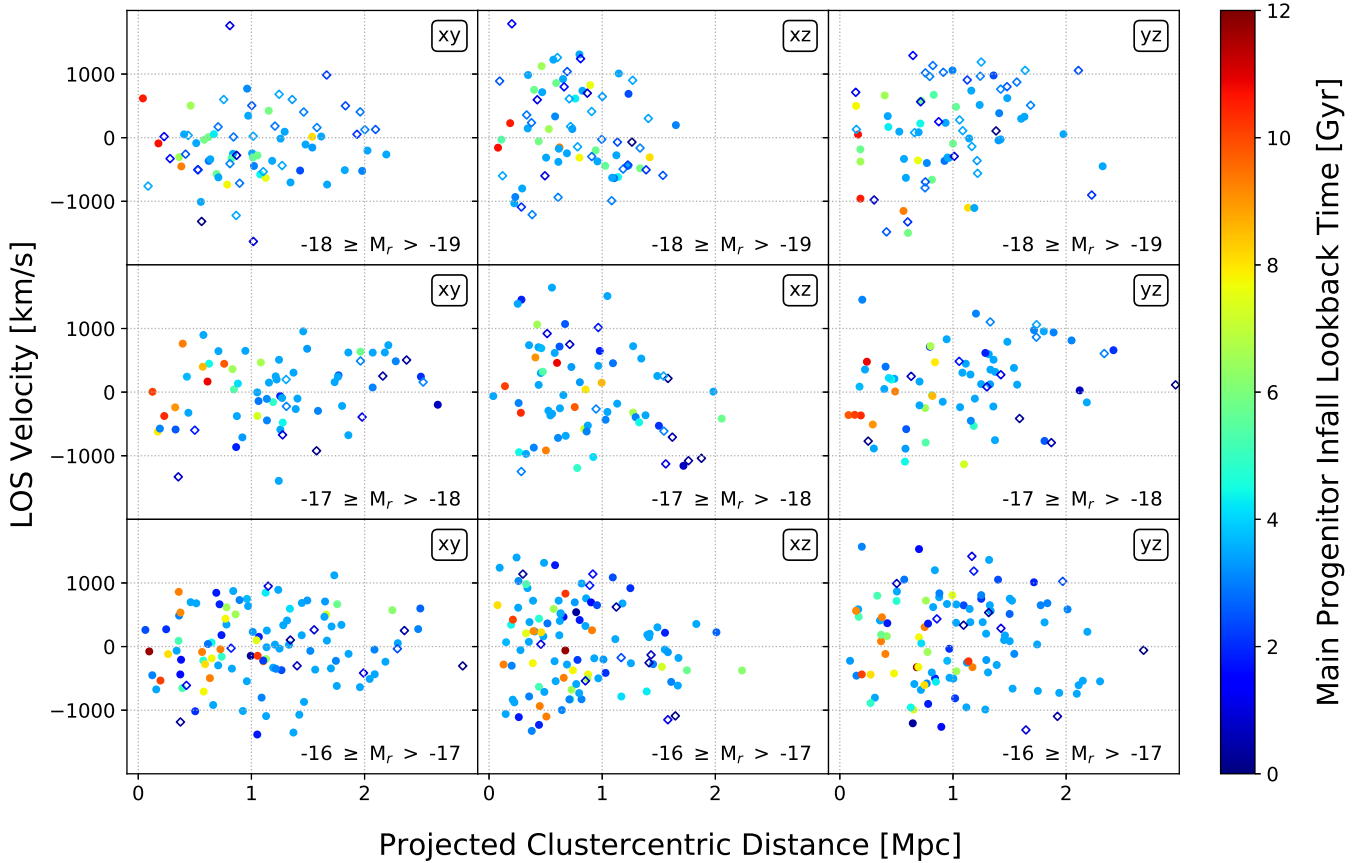}
 \caption{
Observer's phase space for 474 low-mass galaxies with stellar masses between $10^8$ and $10^{10} \mbox{M}_{\odot}$ in the second most massive cluster in the Illustris simulation \citep{Vogelsberger2014,Vogelsberger14b}, shown separately for different intervals of absolute magnitude. Colors indicate the lookback times to the galaxies' infall into one of their host cluster's main progenitors, with red symbols denoting early and blue symbols denoting late infallers. Filled circles represent red sequence galaxies, empty diamonds denote galaxies that are not part of the red sequence.
}
\label{fig:illustris}
\end{figure*}

In the context of cosmological simulations, more investigations of the phase-space distribution of galaxies and its potential signatures of environmental histories would certainly be promising, and would also serve to evaluate possible observational biases. As an example, we present in Fig.~\ref{fig:illustris} the phase-space distribution of dwarf galaxies in the second most massive galaxy cluster ($3.8 \cdot 10^{14} \mbox{M}_{\odot}$) of the Illustris simulation \citep{Nelson15}, which experienced a nearly equal-mass merger just 3.5 Gyr ago. The figure shows three different projections and subdivides the galaxies by the same magnitude intervals as in our previous figures.
Galaxies are color-coded by lookback time to their infall into the main progenitor of the cluster; red-sequence galaxies are represented by filled circles.

There is some clustering and some asymmetry in the phase-space distribution, but it is not as strong as what we see for the Virgo early types in the interval $-17\geq M_r > -18$\,mag.
The differences between the projections are striking: galaxies with different infall times can be much more mixed up in one projection than in the other. In this system, the merger is closest to the $y$ direction, i.e.\ the central column of Fig.~\ref{fig:illustris} corresponds to an observer's line of sight that looks into the merger.
Consequently, we see that the distribution of galaxies in phase space is least relaxed in the central column, particularly in the two most luminous bins.
This illustrates the limited observer's situation, but also the potential of such studies to help interpreting the observed motion and distribution of galaxies in terms of environmental assembly.

%-----------------------------------------------------------------------------------
%-----------------------------------------------------------------------------------
\section{Conclusions}
\label{sec:conclusions}

Virgo is an actively accreting cluster, with several groups in various stages of infall. Virgo galaxies' properties reflect its active infall history, and studying the morphologies, colors, and internal dynamics of these galaxies is useful to understand the build-up of the cluster as well as the environmental history of the galaxies. In this paper, we have simultaneously considered the properties of galaxies themselves and their current dynamical state in the cluster within the cluster's phase space. We find that the properties of galaxies -- their luminosities and morphologies -- are correlated with their position in phase space. Overall, late-type galaxies are found at larger clustercentric radii, as expected for recent infallers. We also find that the distribution of late-types is asymmetric in velocity space; this also holds for early-type dwarfs with magnitudes $-17 \geq M_r \geq -18$. Low-mass late types as well as early-type dwarfs with disk signatures therefore cluster in phase-space. This asymmetric distribution and clustering indicates recent and ongoing infall. The overall diversity in the galaxy properties across phase space therefore points to Virgo's current state of active assembly.

The intermediate magnitude bin of $-17 \geq M_r \geq -18$, which is around where galaxies transition from giants to dwarfs,
has a particularly interesting phase-space distribution of galaxies. Specifically, within a narrow radial and velocity bin near \mbox{$d_{\rm M87}$$\,\approx\,$$1.15$\,Mpc} and \mbox{$v_{\rm rel}$$\,\approx\,$$700$\,km\,s$^{-1}$}, we find an overdensity of early-type dwarfs
that nearly all show signatures of disk components
and have intermediate to young stellar ages compared to the overall population. They are not spatially clustered, but are azimuthally distributed around the center of the cluster. 
In comparing that phase-space distribution of galaxies with predictions from a basic model for group infall, we find evidence for a previously unrecognized group infall event that occurred $2-3$\,Gyr in the past in the Virgo cluster. This infall likely took place along our line of sight producing a radially distributed subset of dwarf cluster members that share common kinematics. This is the first detection of this type of kinematic structure 
{for a previously gravitationally bound group of galaxies, specifically using the dynamics of dwarf galaxies, which are greater in number and therefore provide more robust statistical information than more massive former group members.}

The phase-space properties of these galaxies combined with their sizes, colors, and morphologies can help us understand when and how they have been transformed. Gas-poor galaxies that have stopped forming stars, but still retain disk structures, were likely stripped of their gas by ram pressure, but not morphologically transformed by tidal forces or mergers. If these galaxies are found at large clustercentric radii with high relative velocities, then they are likely recent infallers; they must have been stripped of their gas in the prior group environment. Their structure and internal dynamics probably still reflect an earlier epoch of their evolution.
Comparing their properties to those of early-type dwarfs in galaxy groups, as well as to those closer to the Virgo core, can thus provide insight as to which environment is most critical for determining the galaxies' structure, dynamics, and stellar population properties.

%-----------------------------------------------------------------------------------
%-----------------------------------------------------------------------------------
\section*{Acknowledgments}
We thank Sanjaya Paudel, Soo-Chang Rey, Suk Kim, Glenn van de Ven, Bruno Binggeli, and Helmut Jerjen for helpful discussions and suggestions.

T.L., C.E.\ and L.U.\ acknowledge financial support from the European Union's Horizon 2020 research and innovation programme under the Marie Sk\l{}odowska-Curie grant agreement No.\ 721463 to the SUNDIAL ITN network. C.E.\ is supported by the Deutsche Forschungsgemeinschaft (DFG, German Research Foundation) through project 394551440.
RV was supported by an NSF Astronomy and Astrophysics Postdoctoral Fellowship under award AST-1501374.
JSG thanks the University of Wisconsin Foundation for research support through his Chair Rupple Bascom Professorship.
This research has made use of NASA's Astrophysics Data System Bibliographic Services and of the NASA/IPAC Extragalactic Database (NED) which is operated by the Jet Propulsion Laboratory, California Institute of Technology, under contract with the National Aeronautics and Space Administration.

T.L.\ thanks Caf\'e Fresko Heidelberg for inspiration.
%\bibliography{bluegroup}

% APPENDIX


%\newpage

\appendix

\section{Galaxy images}
Here we show visual images of Virgo galaxies in the $-17 \geq M_r \geq -18$ luminosity bin,  arranged according to their relative positions in the four phase-space regions defined in Figure~\ref{fig:magdistribution}. The highlighted grey box in the upper right hand corner shows the likely remnant of an infalling group -- dominated by early-type galaxies with disks.

\begin{figure*}[!htbp]
\includegraphics[width=\textwidth]{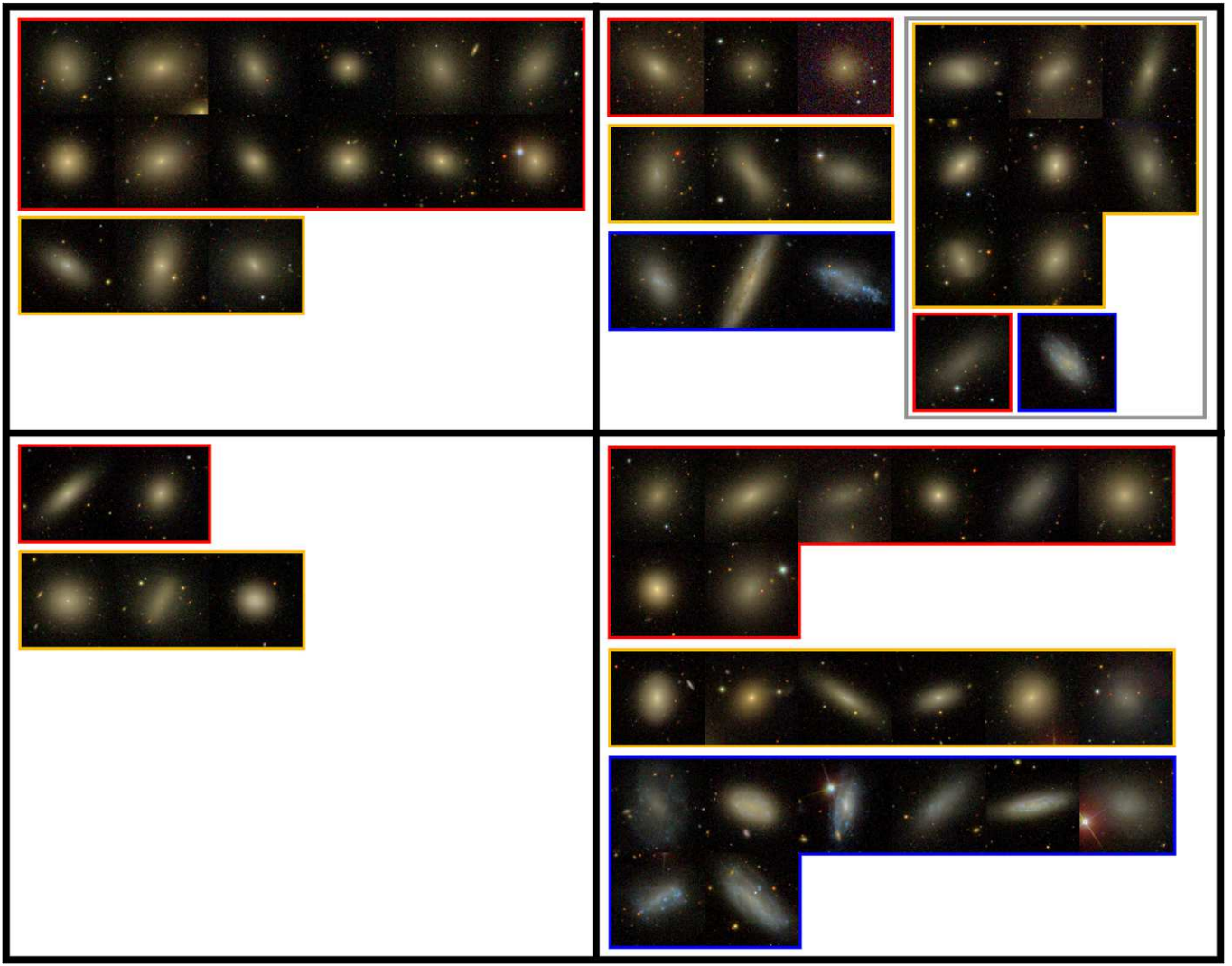}
 \caption{SDSS-IV DR14 \citep{sdssdr14} images of galaxies in the $-17 \geq M_r \geq -18$ luminosity bin, organized by their location in the phase-space diagram. Each image is 100 arcsec wide. The left and right hand side correspond to projected cluster-centric distances of $0-0.75$\,Mpc and $0.75-1.5$\,Mpc. The upper and lower sides are positive and negative radial velocities with respect to the cluster.  Red boxes encompass galaxies classified as ellipticals (early types without disks), blue boxes encompass late-type galaxies, and orange boxes encompass early types with disks. The grey box in the upper right region is the phase-space region with the unique aggregation of early types with disks that is likely the remnant of an infalling group.}
\label{fig:images}
\end{figure*}

\end{document}